# Estimands for Early Phase Dose Optimization Trials in Oncology

**Ayon Mukherjee** *,1, **Jonathan L. Moscovici** 2, and **Zheng Liu** 3

1 Population Health Sciences Institute, Newcastle University, Newcastle, United Kingdom
2 Center for Statistics in Drug Development (CSDD), IQVIA, Montreal, Canada
3 Novella Clinical Full Service, IQVIA, Melbourne, Australia

Phase I dose escalation trials in oncology generally aim to find the maximum tolerated dose (MTD). However, with the advent of molecular targeted therapies and antibody drug conjugates, dose limiting toxicities are less frequently observed, giving rise to the concept of optimal biological dose (OBD), which considers both efficacy and toxicity. The Estimand framework presented in the addendum of the ICH E9(R1) guidelines strengthens the dialogue between different stakeholders by bringing in greater clarity in the clinical trial objectives and by providing alignment between the targeted estimand under consideration and the statistical analysis methods. However, there lacks clarity in implementing this framework in early phase dose optimization studies. This manuscript aims at discussing the Estimand framework for dose optimization trials in oncology considering efficacy and toxicity through utility functions. Such trials should include Pharmacokinetics (PK) data, toxicity data, and efficacy data. Based on these data, the analysis methods used to identify the optimized dose/s are also described. Focusing on optimizing the utility function to estimate the OBD, the population-level summary measure should reflect only the properties used for the estimating this utility function. A detailed strategy recommendation for intercurrent events has been provided using a real-life oncology case study. Key recommendations regarding the estimand attributes include that in a seamless Phase I/II dose optimization trial, the treatment attribute should start when the subject receives the first dose. We argue that such a framework brings in additional clarity to dose optimization trial objectives and strengthens the understanding of the drug under consideration that would enable the correct dose to move to Phase II of clinical development.

## 1 Introduction

Dose-finding trials in oncology are essential to establish recommended doses for later-phase drug development. Traditionally, Phase I trials developing conventional cytotoxic chemotherapy used designs to identify the maximum tolerated dose (MTD), which is typically defined as the highest dose of a drug or treatment that does not cause unacceptable toxicity in a specified proportion of patients. Thus, the MTD represents the most efficacious dose that is safe. This assumption, however, is often questionable for immunotherapy, antibody drug conjugates, biospecific antibodies and small molecule targeted therapies. For these novel therapies, drug efficacy does not necessarily increase monotonically with toxicity as the dose level increases. For example, immune checkpoint inhibitors (ICI) work by reactivating the immune system and, hence, reestablishing its ability to combat tumors. Checkpoint proteins are receptors on immune cells that can be activated to block the immune response, such as checkpoint proteins on T cells (eg, PD-1 and CTLA-4). The ICIs bind to checkpoint receptors in T cells and release "brake" such that T cells can kill cancer cells, achieving treatment efficacy. Once checkpoint binding is saturated, increasing the ICI dose further does not increase treatment efficacy. Moreover, some of these novel therapies demonstrate minimal toxicity even at high doses, making it unlikely that MTD can be achieved. This occurs when the novel compound under investigation has a benign toxicity profile such as the compound NM21-1480 in the

---

*Corresponding author: e-mail: ayon.mukherjee@newcastle.ac.uk

 

Numab Therapeutics AG trial [Luke et al., 2022]. Therefore, the dose-finding design paradigm, which was developed to find the MTD, may not be suitable for such novel therapies.

The Oncology Center of Excellence (OCE) of the U.S Food and Drug Administration (FDA) initiated Project Optimus with the aim of reforming the dose optimization and dose selection process in oncology drug development [Opt, 2024]. In this reform, emphasis is now moving away from identifying the MTD but instead to determining the optimal biologic dose (OBD), which is the dose that maximizes the benefit/risk profile of the compound based on an understanding of the dose and exposure response. While phase I studies assess the toxicity profile of a compound and phase II studies assess its preliminary efficacy, studies that seek to find this balance between the risk and benefit profile of a compound are called seamless phase I/II trials and aim to identify the optimal dose, as well as the schedule, for late-stage (Phase III) assessment within a single clinical trial protocol. Among the different types of seamless Phase I/II trials proposed by Jaki et al. [2023], this manuscript focuses on dose optimization trials with simultaneous evaluation of toxicity and efficacy to identify the OBD as the recommended Phase 2 dose (RP2D).

A few novel model-based, rule-based, and model-assisted designs have been proposed to find the OBD in phase I dose optimization trials by simultaneous evaluation of the probabilities of treatment efficacy and toxicity. Some have been developed using adaptive Bayesian methods to identify a dose level that optimizes patients' risk-benefit trade-off. Building upon the i3 + 3 design for cytotoxic chemotherapies [Liu et al., 2020], the joint i3 + 3 (Ji3 + 3) design was proposed by Lin and Ji [2020]. The Ji3 + 3 design is simple to implement when toxicity and efficacy are both considered as binary endpoints, and allows the monotonic dose-response assumption to be violated. Thall and Cook [2004] introduced the EffTox method, which is an outcome-adaptive, model-based Bayesian procedure that chooses doses of an experimental agent for successive patient cohorts in a trial based on both efficacy and toxicity. Model-assisted designs achieve similar operating characteristics as the model-based designs, however they are as simple to implement in clinical practice as the rule-based designs. Lin and Ji [2021] proposed the Probability Intervals of Toxicity and Efficacy (PRINTE), using toxicity and efficacy jointly in making dosing decisions. The Bayesian optimal interval phase I/II (BOIN12) design [Lin et al., 2020b] is a transparent and flexible trial design to find the OBD that optimizes the risk-benefit trade-off. The adaptation rule in BOIN12 design can be pre-tabulated and included in the clinical trial protocol. Lin et al. [2020b] demonstrated through a simulation study that the BOIN12 method is more robust than the EffTox design because it does not make any model assumptions on the dose toxicity and efficacy curves. It differs from the utility-based BOIN (U-BOIN) design method [Zhou et al., 2019], which is a two-stage design for which the first stage performs dose escalation on the basis of toxicity only, and the second stage uses the toxicity-efficacy trade-off for decision making.

The International Council for Harmonization of Technical Requirements for Pharmaceuticals for Human Use ICH E9(R1) [ich, 2020] defines an estimand as a precise description of the treatment effect reflecting the clinical question posed by the trial objective. It summarizes at a population-level what the outcomes would be in the same patients under different treatment conditions being compared. It is recommended to specify estimands as clearly as possible, providing guidelines and a framework for doing so. According to this estimands framework principle, as depicted in Figure 1, "trial planning should proceed in sequence" so that objectives translate to estimands, which lead to estimators and the estimates [Bell et al., 2021]. The guidance acknowledges that while "the main focus is on randomized clinical trials; the principles are also applicable for single-arm trials and observational studies. Englert et al. [2023] pointed out that despite the pertinence of the ICH E9(R1) guidelines to all types of trial phases or data types, research and applications of the estimands framework have mainly been focused on randomized controlled trials as regulatory interest for implementation of the estimand framework is greater in confirmatory clinical trials. The authors, as a part of the Early Development Estimand Nexus (EDEN) task force, introduced the estimands framework in the single-arm early phase setting to ensure common understanding and consistent definitions for key estimands in early phase oncology trials. While developing the estimands framework, the authors focused on non-randomized oncology Phase 1b expansion studies or Phase 2 studies assuming the MTD to be already established and that the recommended Phase 2 dose (RP2D) is usually defined at the end of the Phase 1b. However, this is usually not true in dose optimization paradigms targeting the OBD during the escalation process in Phase I trials.

This manuscript focuses on extending the work of Englert et al. [2023] by constructing an estimand framework in the dose optimization studies which are single-arm non-randomized studies to identify the OBD through a utility function that considers the patient's risk-benefit trade-off, which presents different requirements from MTD trials. This manuscript focuses on the BOIN12 [Lin et al., 2020b] design methodology while constructing the estimand framework, as it is simple to implement in practice and therefore widely used in real-life clinical trials, especially for developing Chimeric Antigen Receptor (CAR) T cells therapy [Pan, 2024]. However, the proposed framework is applicable for other dose optimization trial designs to identify the OBD through a utility function, for which a case study is also presented.

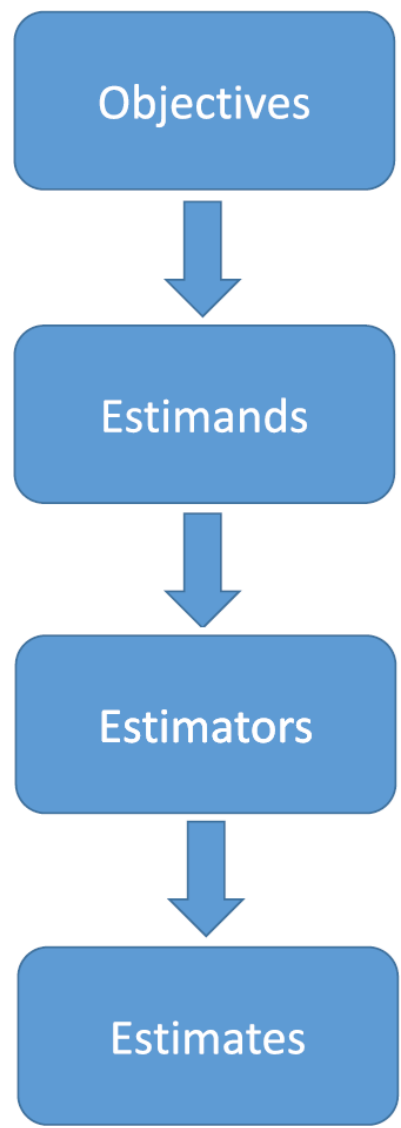

**Figure 1**　　The Central Framework of Estimands According to ICH E9(R1)

## 2  The Estimand Attributes in Dose Optimization Studies

Similarly to a randomized controlled study, it is important to define the attributes of the estimand to be considered for early phase dose optimization studies. In these adaptive trials, the risk-benefit trade-off of patients is optimized by their dose desirability often measured by a utility function that combines the information on safety and efficacy accumulated during the trial process. The primary clinical objective in such trials is to characterize the dose desirability of patients who are being treated by the novel compounds and to identify the OBD, if such a unique OBD exists based on a pre-defined admisibility criteria. An estimand of interest in such trials is the utility function that optimizes the risk-benefit profile of the patients, using which an incoming cohort of patients is assigned to an admissible desirable dose measured by this function and can identify the OBD as the admissible dose with the maximum expected utility at the end of the trial. The attributes below are used to construct the estimand of interest for dose optimization studies and must be clearly specified in a clinical trial protocol. In the event that several estimands are of interest, as mentioned earlier it would be beneficial to consider their attributes jointly to ensure strategies are consistent with available data.

### 2.1  Clinical Trial Objectives

For a seamless phase I/II clinical trial, the primary objective is to characterize the desirability of the dose in patients by optimizing the toxicity and efficacy profiles of the drug candidate, and to identify the OBD, when it exists, or to terminate the trial early if no dose satisfies the desired pre-defined quality requirement. An OBD in a seamless phase I/II clinical trial can be defined based on the clinical question of interest of the trial. In this manuscript, OBD is defined as the dose that has the highest desirability in terms of the risk-benefit trade-off that is measured by a utility function. Regarding toxicity, the selected dose/s during treatment should be well tolerated with minimized dose modifications such as dose interruption, dose reductions, and discontinuations, with few Grade 3/4 and less cumulative Grade 1/2 adverse events of interest. The frequency and impact of such reactions along with duration of interruption should be carefully assessed and considered while selecting the dose level for subsequent clinical trials. For efficacy, it is necessary to determine first if the drug candidate has a clinical effect or not; second, if the drug effect increases with higher doses or exposure; and third, investigate the characteristics of the dose/exposure and effect relationship. Ultimately, the optimized therapeutic dose(s) considering both toxicity and efficacy must be determined.

In dose optimization studies, the primary objective is to identify the OBD, as opposed to finding the MTD in dose escalation trials for cytotoxic chemotherapies. Secondary objectives may include PK goals such as dose characterization [Akacha et al., 2021]. Therefore, the primary estimand for dose optimization trials needs to define the OBD stating the toxicity and the efficacy criteria based on which the optimization would be made. The secondary estimands in such cases can define the PK parameters under consideration for the respective trials based on its secondary objectives. Appropriate strategies need to be implemented to handle each estimands. The dose optimization strategy must however include PK data and a detailed analysis plan to calculate PK parameters to support dose/exposure response analyses for safety and efficacy. Although ICH E9(R1) indicates the importance of defining clinical trial objectives clearly, since they lead directly to the choice of estimands, there is a lack of guidance in specifying the objectives themselves. Care must be taken in the choice of these objectives and typically involve a multi-disciplinary team prioritizing the key stakeholders of the trial [Bell et al., 2021].

### 2.2  Endpoints/ Variable of Interest

The primary endpoint for a seamless phase I/II dose optimization trial with simultaneous evaluation of toxicity and efficacy is defined on a subject level. Generally in such trials, the objective response rate (ORR) assessed by the investigator according to RECIST v1.1 [Eisenhauer et al., 2009] constitutes an efficacy endpoint and the incidence dose limiting toxicities and other safety measures as characterized and tabulated by the latest version of Common Terminology Criteria for Adverse Events (CTCAE) of the National Cancer Institute (NCI) constitute the toxicity endpoint. The toxicity and efficacy measure commonly used to identify OBD that optimizes the risk-benefit trade-off between patients are the dose limiting toxicity rate (DLT) and the response rate, respectively. DLTs are side effects of a compound that are serious enough to prevent an increase in dose or level of that treatment. This may also be accompanied by safety endpoints such as the toxicity measures induced by the compound under investigation. The toxicity endpoints are usually the number of Grade 3/4 severe adverse events, mild Grade 1/2 symptomatic adverse events, dose modifications as well as patient-reported outcome data. In late phase trials, the efficacy endpoints are overall response rate, duration of response etc. In early phase trials however, the efficacy endpoints can be limited to various biomarkers and surrogates, but often also include progression-free survival (PFS), ORR, overall survival (OS), duration of response (DOR) as well, which can also be predictive of late phase trial endpoints. The endpoints are grouped into continuous variable, categorical variable (e.g. responder vs. non-responder) and time-to-event variable (e.g. death). If responses are delayed, some observed or latent pharmacodynamic (PD) biomarker can be used as a surrogate of the actual clinical response. Exposure-response and PK data can also define the changes in dose and dosing regimens that account for intrinsic and extrinsic patient factors. Both the magnitude of an effect and the time course of effect are important to choose dose, dosing interval, and monitoring procedures, and even to deciding what dosage form (e.g. controlled-release dosage form) to develop. PK parameters may also be measured across single or multiple doses, as the data permits and every trial protocol should be be accompanied by a PK sampling and analysis plan.

### 2.3 Target Population

Dose optimization trials typically target a patient population with a specific oncology indication and a therapeutic line. Often this consists of patients with advanced or metastatic malignancies who have shown disease progression on at least one or two prior lines of therapy. They can also consist of patients who have experienced relapse from or are refractory to a previous treatment.

There is a close connection between the target subject population and the clinical trial objective which is motivated by the clinical question of interest for the study. While the subject analysis set is the sample of patients analyzed in the clinical trial, it is important to understand and differentiate the target population from the analysis set. However, it is critical to be mindful that the analysis set represents the target population of interest. To assess anti-tumor activity in the trial, the analysis set typically consists of patients who have received at least one dose of the treatment under consideration.

### 2.4 Treatment Under Consideration

The estimand attribute of treatment condition for a dose optimization trial should specify the given compound, combination of compounds, or a sequence of the compounds, which are administered during the optimization schedule. It should specify the competing dose levels under consideration in the trial. The planned dosing regimens for the compound or the combination, along with the frequency in which the doses would be administered, need to be carefully chosen at the design stage based on historical information of related clinical data or the pre-clinical study, and need to be included as part of the treatment condition description [Francois Mercier and Victor, 2024].

In randomized treatment comparisons, the treatment condition is normally considered as starting at the time of randomization [Englert et al., 2023]. For dose optimization trials, the equivalent would be when the patient has passed the screening period and is eligible for receiving the planned dose. As there is a significant time gap between collection, analysis, and evaluation of the sample for various measurements during the screening period, precisely specifying the date when all screening activities are completed is an important requirement. Similar to the recommendation of Englert et al. [2023] for Phase 1b/Phase 2 trials, using the time of the first dose as the landmark to define the start of the "treatment condition of interest" also applies for dose optimization trials. The date of first dose can be easily collected and precisely determined for each subject. In most dose optimization studies involving novel therapies such as checkpoint inhibitors and chimeric antigen receptor T-cell (CAR-T) therapies there is a significant gap between the intent-to-treat and first dose. The treatment period for the assessment of subjects in the dose optimization study of such novel therapies would effectively start after the actual initiation of treatment and not after the intention of treatment. The definition of the treatment condition of interest should also specify if any assessments after the protocol-defined end of the treatment period will be taken into account [Eisenhauer et al., 2009], or if the treatment would be considered at the expansion stage of the trial such as in Phase Ib.

### 2.5 Population-Level Summary Measure

BOIN12 [Lin et al., 2020b] is a model-assisted adaptive design that accounts for the risk-benefit trade-off of the patients in the trial by identifying the dose with the highest desirability using a utility measure that combines toxicity and efficacy outcomes collected during the trial. The estimate of the OBD is the mode of the estimated dose-desirability curve. These utility scores allow for direct comparison of doses as information is collected and doses are escalated/de-escalated accordingly in order to search for the OBD. The utility scores can combine information regarding safety and efficacy, such as binary outcomes of toxicity/no toxicity and efficacy/no efficacy for each subject and produce a score describing the utility of that outcome. Ideally an outcome of efficacy and no toxicity produces the highest utility, but elicitation of expert information is required to determine the utility there might be in a efficacy/toxicity outcome as well as a no-efficacy/no-toxicity outcome for the indication and patient population of interest. The study-defined estimand aligned with this objective would also be focusing on the estimation of an absolute utility score based on which the next patient would be assigned to the dose which is most desirable considering the toxicity-efficacy trade-off. For conventional binary toxicity and efficacy outcomes (such as DLT rates for toxicity and response rates for efficacy), defining the toxicity-efficacy trade-off directly based on the marginal efficacy probability and the marginal toxicity probability can provide an intuitive interpretation to the utility function to be optimized to identify the OBD. As an example of a utility framework, the respective utilities for each combination of safety/efficacy binary outcomes can be expressed by $Y$, $Y_e$ and $Y_t$ as follows. Set $Y_e = 0$ if no efficacy was observed, and $Y_e = 1$ if efficacy was observed. Set $Y_t = 0$ if no toxicity (such as Dose Limiting Toxicity) was observed, and $Y_t = 1$ if toxicity was observed. For example, for Y, set $Y = 1$ if $(Y_e, Y_t) = (0, 1)$; $Y = 2$, if $(Y_e, Y_t) = (0, 0)$; $Y = 3$, if $(Y_e, Y_t) = (1, 1)$; and $Y = 4$, if $(Y_e, Y_t) = (1, 0)$., where $Y_e$ in a binary indicator for efficacy, such as complete or partial response, and $Y_t$ is a binary indicator for toxicity, such as Dose Limiting Toxicity (DLT). As mentioned in Lin et al. [2020b], the BOIN12 design can be generalized to multilevel or multiple outcomes as well. $Y = 1$ indicates the least favorable clinical outcome, and Y=4 denotes the most favorable clinical outcome. As described by Zhou et al. [2019], define $\pi_{jk} = Pr(Y = k | d = j)$, $k = 1, ..., K$ and $j = 1, ..., J$, with $\sum_{k=1}^{K} \pi_{jk} = 1$ where $d$ denotes the dose level and $K$ is the number of possible outcomes for $Y$ ($K = 4$ in our example here, but can be generalized to more). It is assumed that $Y$, at each dose level, follows a Dirichlet-multinomial model where:

$$Y = k | d = j \sim Multinomial(\pi_{j1}, ..., \pi_{jK}).$$

$$(\pi_{j1}, ..., \pi_{jK}) \sim Dirichlet(a_1, ..., a_k)$$

where $a_1, ..., a_k > 0$ are hyperparameters. Setting $a_k = 1/K$, $k = 1, ..., K$, the prior is vague and equivalent to a prior sample size of 1. Once an interim decision time is reached, a posterior distribution can be calculated. Assuming that $n_j$ subjects have been treated as dose $d = j$, where $n_{jk}$ subjects experiences outcome $Y = k$, where $n_j = \sum_{k=1}^{K} n_{jk}$. Given the observed data $D_j = (n_{j1}, ..., n_{jK})$, the posterior distribution of $\pi_j = (\pi_{j1}, ..., \pi_{jK})$ is

$$\pi_j | D_j \sim Dirichlet(a_1 + n_{j1}, ..., a_K + n_{jK})$$

The desirability of the doses are measured using a utility function. Call $\psi_k$ the utility assigned to the outcome $Y = k$, where $k = 1, ..., 4$ in our case. $\psi_k$ should be elicited from expert clinical and study team members' opinion to indicate the risk-benefit for medical treatments. A guideline for this would be:

1. Assign $Y = 1$ as $\psi_1 = 0$ as the least desirable outcome (toxicity, no efficacy), and $\psi_4 = 100$ for $Y = 4$ as the most desirable outcome (no toxicity, efficacy).

2. Ask the clinical expert in the study team to use these two reference utilities as a guide to assign utility scores for $\psi_2$ (no toxicity, no efficacy) and $\psi_3$ (toxicity, efficacy).

Clinical experts with experience in the target indication would be able to provide expert input on the compound under investigation, whether toxicity can be tolerated for additional efficacy and provide consensus on the relative utility scores. An example could be

$$\psi_1 = 0, \; \psi_2 = 10, \; \psi_3 = 60, \; \psi_4 = 100$$

Where 100 (maximum utility) is assigned to $\psi_1$ (no toxicity, efficacy), 60 is assigned to $\psi_3$ (toxicity, efficacy), 10 (lower utility) is assigned to $\psi_2$ (no toxicity, no efficacy), and 0 (lowest utility) is assigned to $\psi_4$ (toxicity, no efficacy). In this illustration, the relative utility score rewards the response (ie, $Y_e = 1$) more, in the presence of toxicity (ie, $Y_t = 1$), by assigning a larger value to $\psi_3$ (60 versus 10) than $\psi_2$. This is appropriate for a clinical trial where toxicity for the compound under investigation can be well managed and response is highly desirable (eg, leading to long survival). Given the values of $\psi_k$, the mean utility for dose $j$ is computed the weighted average of $\pi$:

$$U_j = \sum_{k=1}^{K} \psi_k \pi_{jk}.$$

In most real-life early phase oncology trials, $Y_t$ an $Y_e$ are considered to be binary rate outcomes. In such cases, another common approach to define the toxicity-efficacy trade-off is directly based on the marginal efficacy probability $\pi_{e,j} = Pr(Y_e = 1 | d = j)$ and the marginal toxicity probability $\pi_{t,j} = Pr(Y_t = 1 | d = j)$, which can be expressed as

$$U_j^M = \pi_{e,j} - w.\pi_{t,j}.$$

where $w$ is a pre-specified weight. This represents that patients are willing to trade an increase of $w$ in the DLT rate for a unit increase in the efficacy response rate. If $w = 0$, we obtain the special case that the dose with the highest efficacy is the most desirable. Theorem 1 of [Zhou et al., 2019] showed that this marginal-probability-based approach $U_j^M$ is a special case of the mean utility approach $U_j$.

To ensure patients are not assigned a toxic and/or futile dose, two dose-admissibility criteria are used by the BOIN12 design to decide which doses can be used to treat patients. Given the interim data $D = D_j$, for a clinician-specified toxicity upper-limit $\phi_t$ and efficacy lower-limit $\phi_e$, dose $j$ is inadmissible, if it meets either one or both of the following two criteria:

$$\text{(Toxic)} \quad Pr(\pi_{t,j} > \phi_t | D) > \delta_t, \quad \text{(Futile)} \quad Pr(\pi_{e,j} > \phi_e | D) > \delta_e,$$

where $\delta_t$ and $\delta_e$ are probability cutoffs that should be calibrated using simulation to ensure desirable operating characteristics. Generally, $\phi_e$ can take the value of the target response rate specified for a standard phase II trial, and the value of $\phi_t$ should be set slightly higher than the target toxicity rate used in conventional toxicity-based phase I designs. The maximum tolerated dose level here is defined as the dose level that has the isotonically estimated toxicity probability closest to the toxicity upper-limit $\phi_t$. If A denotes the set of admissible doses, at the end of the trial process, the OBD is defined as the dose that is admissible and has the highest utility value given by;

$$\text{OBD} = \text{argmax}_{j \in A}(U_j),$$

but not higher than the identified maximum tolerated dose level.

The BOIN12 design deterministically assigns the next cohort of patients to the dose $j \in A$ that has the largest posterior mean utility. However, with this approach, there is a risk of the process staying at at a local suboptimal dose, due to a large variation caused by a small sample size in such early phase trials. An incoming cohort of patients can also be adaptively randomized the to dose $j \in A$, with probability $\omega_j$ proportional to its posterior mean utility, ie,

$$\omega_j = \frac{U_j}{\sum_{j \in A} U_j}$$

As pointed out by [Zhou et al., 2019], the adaptive randomization method can reduce the risk of remaining at a suboptimal dose, but as a trade-off, it tends to treat fewer patients at the OBD. Another approach is equal randomization, where the next cohort of patients is assigned to the set of admissible doses with equal probability. However, it was shown by [Zhou et al., 2019] that none of the methods dominates the others and therefore we would consider for simplicity that the incoming cohort of patients is deterministically assigned to the dose $j \in A$ that has the largest posterior mean utility.

Further discussion can be found in discussions of dose optimization methods such as BOIN12 [Lin et al., 2020b] and UBOIN[Zhou et al., 2019]. In the case of BOIN12, the multinomial method described above can be computationally improved by employing a quasi-binomial method described in the supplementary data of the BOIN12 paper [Lin et al., 2020b]. Expert opinion may influence the values of $\psi_2$ and $\psi_3$ heavily, since the utility and desirability between toxicity/efficacy can depend on the clinical circumstances.

In general, dose optimization methods for seamless phase I/II trials such as BOIN12 [Lin et al., 2020b], UBOIN[Zhou et al., 2019], BOIN-ET [Takeda et al., 2018], EffTox [Thall and Cook, 2004], PRINTE [Lin and Ji, 2021] and others uses measures based on the background statistical model each method is based on, to evaluate and summarize the toxicity/efficacy trade-off values that is the population-level summary measure of interest. As will be explained in further sections, a main difference between the estimand framework for MTD trials and estimands for OBD trials (dose optimisation) could be in the handling of treatment discontinuations due to toxicity. As the targeted therapies have wider therapeutic indices, patients may receive these therapies for much longer periods, potentially leading to persistent symptomatic toxicities, which can be challenging to tolerate over time. If handled using a composite variable approach (handled as failure), a discontinuation due to toxicity would yield the lowest possible utility score.

## 2.6 Intercurrent Events

Intercurrent events (ICEs) refer to medical events or conditions that occur in a study subject during the course of a trial. ICH E9 (R1) defines ICEs as "events occurring after treatment initiation that affect either the interpretation or the existence of the measurements associated with the clinical question of interest". In oncology, intercurrent events can include discontinuation due to adverse events, use of additional anti-cancer therapy (or rescue medication more generally), death, disease progression, adherence to planned treatment regime, surgery, etc. Each potential intercurrent event must be associated with an estimand strategy in order to pre-specify analysis when each type of intercurrent event is observed. This reduces bias and ensures alignment between study goals and analyses. The classification of an event as being intercurrent and the strategies to handle it depend on the clinical question of interest that relates to the study objectives and endpoints of interest. For example, a study drug discontinuation might not be considered as an ICE if it occurs after a complete response is recorded, if the endpoint of interest was indeed to observe a complete response during the treatment period. In this case, the endpoint of interest was observed before the study drug discontinuation, which would disqualify the event from being an ICE. Thus, not all study drug discontinuations (or other similar events) are necessarily ICEs. They need to be considered with the endpoint of interest and the clinical question in the trial to determine if they qualify as ICEs. In an oncology dose optimization study, a potential endpoint of interest could be to observe at least a partial response during the treatment period with no toxicities. If a treatment policy approach is taken, observations after treatment discontinuation due to toxicity would still be considered and if a response is observed both the response and the toxicity will be included in utility/desirability calculations for that dose level. Under a while-on-treatment or composite approach data after the treatment discontinuation would not be considered, as illustrated in Figure 2.

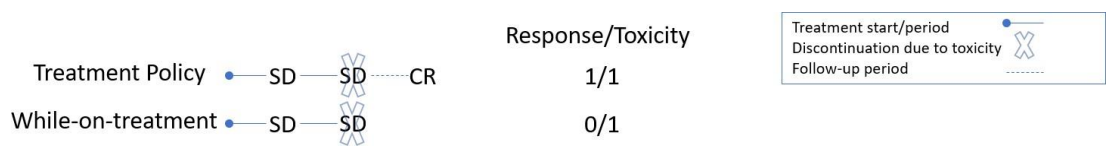

**Figure 2** Illustration of potential ICE handling strategies for treatment discontinuation due to toxicities for dose optimization designs. In both cases, stable disease (SD) is observed at the time of discontinuation, but the treatment policy approach allows inclusion of a complete response (CR) observation during the follow-up period after treatment discontinuation.

However, a while-on-treatment approach would include any observations that occur at the same time as a treatment discontinuation, but a composite approach might not, depending on the composite variable strategy that is employed. As previously mentioned, common intercurrent events also include the use of additional anti-cancer therapy, death, disease progression, adherence to planned treatment regime and surgery. As noted by Englert et al. [2023], although some ICEs may take place around the same time and even be caused by one another (e.g. treatment discontinuation due to toxicity and use of additional anti-cancer therapy), they should be handled separately and clearly defined in the protocol. In general, although every effort should be made to that effect, it must be noted that the list of pre-specified intercurrent events (ICEs) is never exhaustive and it may not be possible to pre-specify all potential ICEs for a trial. If a new type of ICE should emerge, consideration would need to be paid to whether an existing handling strategy could suffice, or if a new one might be needed. Sensitivity analyses could be warranted and amendments to the original plan may be justified. In all cases, thoughtful consideration and a suitable cross-functional collaboration within the clinical trial team would be required to provide some guidance to be included in the protocol on how to handle the cases of potentially emerging ICEs those are resembling but not falling into one of the pre-specified categories. A discussion of the most common types of ICEs in oncology dose optimization studies and recommended handling strategies is presented in the next sections.

he and in practice there should be

## 3 Strategies for Handling Intercurrent Events

Various strategies can be considered for handling intercurrent events for dose optimization studies. A clinical trial may contain several estimands of interest related to its respective trial objectives, and the intercurrent events need to be handled by appropriate strategies for each estimand. If patients experience overlapping events and, if these events are handled using different strategies, the order of application for these strategies must also be specified and will depend on the clinical context [Polverejan et al., 2023]. Each estimand under consideration is a direct consequence of the trial objective it is targeting [Bell et al., 2021]. The handling strategies of the intercurrent events (ICEs) relate to the clinical question of interest that is being answered during the trial, and this is related to the trial objectives. ICE handling strategies may differ during the dose escalation stage and at the final analysis stage, when the final decision on OBD are made because the clinical context of the dose escalation stage and the final analysis stage can be different. A trial investigator may want to achieve a quick escalation process based on the emerging data to make the trial short and move the molecule to the further process of the drug development, whereas at the final stage they would want to make a robust prediction of the OBD based on all the accumulated data and identify the recommended phase II dose (RP2D) that should move ahead in the developmental process. A trial might use a treatment policy strategy by ignoring subsequent events during the dose escalation stage to get a quick, albeit potentially biased, view of impact of the estimated desirable dose for the next cohort during the escalation stage, while the final analysis might employ a more sophisticated strategy like a "principal stratum" or "hypothetical" strategy" that accounts for treatment discontinuations or other intercurrent events more precisely. For the hypothetical strategy, the OBD is defined under the hypothetical scenario in which the intercurrent event is prevented. It must be ensured that decisions underlying hypothetical estimands should not be made deterministically. It must ensure whether estimating a hypothetical estimand is reasonable, and what data should be used in the analysis.

All considerations in this section apply to the utility function of a dose optimization study which combines toxicity and efficacy to search for an optimal dose level. This utility function defines the desirability score of a dose level and an incoming cohort of patients is assigned to the dose level which corresponds to the mode of this desirability curve across the competing dose levels. For ease of exposition, we assume that toxicity and efficacy are represented by binary endpoints, that are observed instantaneously. In some immunotherapy trials, such assumptions though can be fairly stringent as toxicity and/or efficacy may be late onset, causing logistical difficulties while implementing the dose optimization trial design. Under such circumstances, the missing data imputation method can be used in the time-to-event design setting for dose optimization trials to facilitate the real-time decision making of dose escalation/de- escalation. The choice of intercurrent event handling strategies affects the desirability scores computed and thus the choice of OBD. These choices may differ from the MTD case, and must be considered carefully. Sensitivity analyses could be used as needed to examine the robustness of analysis results to various handling methods.

### 3.1 Discontinuing Treatment due to Toxicities

#### 3.1.1 Dose utility regardless of treatment discontinuation due to any cause

In early phase dose escalation studies, patients often discontinue treatment due to excessive dose limiting toxicities (DLTs). Patients may receive targeted therapies for much longer periods than the cytotoxic chemotherapies, potentially leading to persistent low-grade symptomatic toxicities, which can be challenging to tolerate over time. Traditional dose escalation designs targeting MTDs often do not adequately evaluate such low-grade symptomatic toxicities, dosage modifications, drug activity, dose and exposure-response relationships, and relevant specific populations. In a clinical trial implementing a dose optimization strategy considering a patient's desirability score, the treatment discontinuation due to such high-grade or low-grade symptomatic toxicities often makes the question of interest to be about the utility of a dose level regardless of the cause of treatment discontinuation. This is because the desirability score is measured through a utility function of the drug efficacy and toxicity measures.

This dose optimization question is addressed by a treatment policy strategy. Namely, follow-up assessments need to be performed, and response assessments for measuring dose desirability would have to be collected beyond treatment discontinuation. The method of collection of the response assessments may be unplanned depending on sponsor decision. Some sponsors prefer to target the desirability during the DLT period, reflected by a "while-on-treatment" strategy. This only considers the outcomes of any safety or efficacy information collected up to the time of the treatment discontinuation to obtain the desirability score.

The particular method used depends on the design being used for dose optimization. For instance, when implementing a BOIN12 design strategy, calculation of dose desirability can be pre-tabulated and included in the trial protocol before the trial starts using the quasi-beta-binomial model which converts complex desirability calculations into simple beta-binomial modeling [Lin et al., 2020b]. A while-on-treatment strategy would involve data to be included in the quasi-beta-binomial model prior to treatment discontinuation. As pointed out by Englert et al. [2023], the while-on-treatment strategy impacts the definition of the variable, as the observation time of interest is restricted to the time before the treatment discontinuation.

### 3.1.2 Optimal drug dose if the drug were tolerable for all patients

In first-in-human studies, the drug formulation might not be the final one taken forward for licensing. A compound under consideration might ultimately have a benign toxicity profile which may be unknown to the sponsor prior to the start of the trial, since it may only be discovered in further formulations of the drug that potentially remove toxicity responses without compromising efficacy. Examples of such cases could be changing from immediate-release to extended-release oral formulations which could improve safety by minimizing peak exposure and increase patient compliance by requiring a less frequent dosing regimen, or by reducing degradation/sensitivity due to the storage environment and by reducing impurities in manufacturing. In such cases, a relevant question to address would be: What would be the optimal dose level if all the patients were able to tolerate the drug (i.e. if the compound would have a benign toxicity profile)? Though an answer to this does not provide actionable information and is unlikely to be relevant for regulatory purposes, it might be useful in selecting a more informed dose level to the next phase of the drug development process to enhance the success probability in efficacy comparison at later phases. The optimum dose here is determined based on the shape of the efficacy curve across the dose range guided by the dose-response relationship. To address this question, the focus of the analysis would be on the subpopulation of patients who would be able to tolerate and continue taking the experimental treatment as planned. Analysis in this manner would correspond to a principal stratum strategy. For dose optimization trials escalating dose of a single compound as a monotherapy or in combination, there are two principal strata: patients who would experience the toxicity leading to treatment discontinuation and patients who would not discontinue the treatment due to toxicities. As pointed out by Englert et al. [2023], a subgroup analysis differs from a principal stratum strategy. The mode of action of the treatment, pre-clinical or external related study data would need to be used to identify these strata from baseline characteristics. With a dose optimization trial, any modeling of what might occur in an alternative world may lack convincing data on which to base a statistical model. Historical data would lack credibility as a basis in such scenarios. In practice, the results from principal stratification can be used to understand causal effects within specific subgroups. However the assumptions required to implement a principal stratum approach for accurate causal inference are quite strong and may be difficult to justify. For example, it is difficult to determine which patients belong to the principal stratum population when they are assigned treatment, since this would require knowing their future intercurrent event status under each treatment strategy [Kahan et al., 2024]. These assumptions are important for identifying and estimating the OBD within the principal strata. Regulators would require strong evidence for such an approach and why the trial's particular characteristics or objectives could lead to a benefit from this method given the assumptions required. If there is a lack of convincing data to identify the strata for implementing a principal stratum approach, a hypothetical strategy would be ideal to handle such a scenario which analyzes data envisaging if the non-tolerators of a dose actually tolerated the optimum dose, however as mentioned in ICH E9(R1), the clinical and regulatory interest of such hypotheticals is limited and would usually depend on a clear understanding of why and how the intercurrent event or its consequences would be expected to be different in clinical practice than in the clinical trial. As a further clarification, the goal of this strategy would not to be to predict which patients would or would not tolerate a drug upon prescription, but rather to address the specific question of what dose would be optimal in a circumstance where all patients tolerated the drug (for example, if the safety formulation of the drug were improved).

A scenario can also be envisaged where patients who experienced toxicities pre-defined as leading to treatment discontinuation would have continued with the treatment. The outcomes here would have to be hypothesized and imputed. Although this scenario may not be admissible from a regulatory perspective, the sponsor may be interested in it during their exploratory stage for internal decision making about the development of the compound. If the dose-response relationship of a drug is quite flat in a hypothetical scenario where all patients would have tolerated the drug, the mode of the desirability curve might not be achieved. In such circumstances the highest dose would be deemed to be selected to move to the Phase II of the development process or it might be decided that it is not worth pursuing the drug further.

### 3.2 Use of Additional anti-cancer therapy

The event of patients taking additional or subsequent anti-cancer therapy impacts the efficacy outcomes and needs to be taken into account in the interpretation of the OBD. In case of response during the period when a patient receives an additional anti-cancer therapy, it may be challenging to ascertain if the observed response is attributed to the investigational treatment, the additional therapy, or to the combination of these. A detailed account on appropriate estimands has been provided by Manitz et al. [2022], in the presence of treatment switching or subsequent anti-cancer therapy in late stage studies with time-to-event endpoints. For seamless Phase I/II dose optimization study with simultaneous evaluation of toxicity and efficacy in oncology, the taking of additional/subsequent anti-cancer therapy may be assumed to be associated with a non-favourable response. For such an assumption, a while-on-treatment strategy could be applied, only considering response to treatment prior to the additional anti-cancer therapy. When implementing such while-on-treatment approach, some summary statistic of the PK endpoint such as the Area Under the Curve (AUC) that tends to be less favourable when based only on outcomes while on treatment can be reported in the clinical study report. The mean based only on outcomes while on treatment can also be considered in such cases, however one should be careful that such a measure of average conceals the non-favourable occurrence of events. In case of assuming a non-favorable response or undesirable toxicity to the treatment under investigation, then a composite variable strategy could be used where patients who take subsequent anti-cancer therapies are considered treatment failures or considered as a part of a toxicity event, which would lower the desirability score.

If the interest lies in estimating the OBD irrespective of the additional anti-cancer therapy taken or alongside the subsequent therapy, then the clinical trial protocol should outline the data collection requirements post-initiation of anti-cancer therapy and a treatment policy strategy can be applied retaining all outcomes and patients, ignoring the presence of such intercurrent events. In this case, the desirability score would have the potential to be higher than in the composite variable case mentioned above. The treatment condition attribute of the concerned estimand under would also be updated so it considers the prescribed treatment and any additional treatments.

### 3.3 Occurrence of Human Anti-Chimeric Antibodies

Dose optimization trials are frequently used for estimating an OBD for an immunotherapy such as chimeric antigen receptor T-cell therapy. For these novel therapies, although toxicity may typically increase with the dose, efficacy may plateau or even decrease at high doses [Lin et al., 2020b]. Such trials in immunotherapy can often provoke an unwanted humoral immune response that is the formation of antidrug antibodies (ADAs), also known as human anti-chimeric antibodies. In cancer immunotherapy, the generation of ADAs is an increasingly important concern as it may explain the failure of certain therapeutic proteins [Smith et al., 2016] and could decrease or suppress the drug exposure.

Neutralizing the ADAs becomes important if its impact on the drug effect is established. As pointed out by Englert et al. [2023] for single-arm phase 1b trials, considering the occurrence of neutralizing ADA as an intercurrent event will depend on additional assumptions made, such as the relationship between the ADAs and the absence of efficacy when the efficacy itself can be difficult to establish due to having longer windows. Therefore, such assumptions are very hard to justify.

The impact of an ADA on drug efficacy and on the dose-response relationship needs to be accounted for appropriately in the trial conduct. Using the hypothetical strategy envisaging the scenario where the treatment would not induce ADAs is one way to account for this if there is a possibility of preventing the future occurrence of the ADAs. This would result in estimating the dose-response relationship in the theoretical scenario where the treatment would not induce ADAs identifying the dose efficacy if ADAs could be prevented in the future. However, this strategy is difficult to justify in a real clinical trial setting aiming for further development. In a clinical study aiming at further development of the compound, the treatment policy strategy, assessing the dose-response relationship with all its characteristics including ADAs, would be a suitable choice for estimating the OBD in early oncology trials.

### 3.4 Death

Terminal events such as death can occur in first-in-human studies specifically for compounds with high toxicity profiles that cannot be well tolerated by patients. Therefore, if a patient dies before the toxicity or efficacy response is observed, it is reasonable to consider the patient as a non-responder for predicting the dose-response relationship and consider that as an event to predict the toxicity probability for that cohort. In such cases, death is considered in itself to be informative about the patient's outcome and is therefore incorporated into the definition of the variable used to estimate the OBD. For binary and multinomial toxicity and efficacy profiles, this would result in a composite variable strategy. Following Englert et al. [2023] as in phase 1b single arm trials, it is also reasonable to consider that even if a positive radiological response is observed at the same time as the occurrence of death the patient should still be considered as a non-responder at the time of death. A component of the clinical question of interest would then be: "What is the dose-response relationship of the compound if, at the time of subject death, failure to achieve the efficacy response is assumed and the dose is considered toxic enough to cause the terminal event?" A composite variable strategy is applied to death, including death in the variable definition as a non-responder and that as the dose limiting toxicity too. This is a conservative choice which lowers the desirability score accordingly.

In certain clinical settings, early death might be unavoidable–but still unpredictable–and is not necessarily an indicator of treatment failure. An element of the clinical question of interest is then "what is the dose-response relationship in a population that does not experience early death?". In such cases, defining a principal stratum and focusing the analysis on patients that would not experience early death is not possible. In this case it is more appropriate to perform a pre-planned sub-group analysis excluding subjects with early deaths, defining a supportive estimand whose population is stated to consist of the subgroup.

In first-in-human oncology trials, deaths are usually pooled together as all-cause mortality. When death is viewed as contributing to the assessment of the success of the treatment in a way that can be represented quantitatively, death could be included in the variable definition using composite variable strategy. For deaths occurring in the trial for causes assessed to be temporary in nature, and unrelated to the study treatment (eg: poliomyelitis), a hypothetical strategy can be incorporated envisaging a scenario in which such deaths would not occur. However, in practice this may be challenging to implement, since it may be difficult to determine the exact cause of death. For sparsely populated dose optimization trials in oncology, even isolated death events may influence the results and interpretation of the study estimand and therefore its impact should be carefully assessed when planning the study.

### 3.5 Adherence to the Planned Treatment Regime

In an early phase dose optimization oncology trial, patients may not fully adhere to the planned medication schedule. This might be because of patients experiencing dose omission, dose interruption, dose reduction, medication error, excessive toxicity at the dose administered during the trial or an excessively delayed efficacy window at a certain dose level. This is common in immunotherapy trials where the toxicity and/or efficacy may be late onset [Lin et al., 2020a], or the planned therapeutic window is very large, causing logistical difficulties when implementing some the adaptive design methods for seamless Phase I/II dose optimization trials. This lack of adherence could have an impact on the trial results as where a patient would not achieve a response, it would be unknown if the response could have been different had they been fully compliant with the planned treatment schedule. While waiting for the complete efficacy evaluation can significantly increase the trial duration, there are approaches [Cheung and Chappell, 2000], allowing only partial evaluation of the response through missing data imputation [Barnett H et al., 2022]. However, patients switching to a different dose level during the trial process could significantly impact the trial results of identifying the OBD. For such an intercurrent event, it would not be known whether the outcome could have been different had they been fully compliant with the planned dosing schedule. Therefore, appropriate strategies are warranted to address such intercurrent events, if deemed helpful in addressing the clinical question of interest.

If adherence to the planned dose escalation method is an expected issue in clinical practice, the treatment policy strategy is recommended. Equivalently, the treatment condition attribute of the concerned estimand could be defined so it is understood as the prescribed dose including potential dose omissions, dose interruptions, dose reductions, dose switching or medication errors. Per ICH E9(R1), subjects should be followed-up and assessed regardless of adherence to the planned course of treatment and that those assessments should be used in the analysis. Appropriate sensitivity analysis needs to be suggested in the clinical trial protocol to explore the assumptions made during the trial design. However, if the question of interest is "what would the OBD be should a patient be adherent with protocol defined dosing plan?," then similar considerations to "discontinuation due to toxicities" would apply, though different strategies may be chosen (e.g. treatment policy, composite, hypothetical, etc.) depending on the questions of interest. It can be useful to handle treatment adherence through separate ICEs and strategies, as related to the question of interest. One consideration could be that clinical trials may collect data under circumstances different from usual clinical practice. In this way, adherence might be very different. Clinical experience can help determine to what degree this occurs and whether it should be accounted for. The proposed treatment policy strategy approach here results from the adaptive dose escalation method for using the status of the dosing compliance as per the study design approach along with the observed outcomes to estimate the OBD. This could raise or lower the desirability score, depending on the results originating from the lack of adherence.

### 3.6 Surgery

As pointed out by Englert et al. [2023], in trials focusing on target populations with unresectable tumors, a patient enters an early phase clinical trial under the auspices that they have inoperable cancer. However, the tumor may become operable during their treatment process, while still not meeting the response criteria. The investigator may decide to surgically remove target and/or non-target lesions during the treatment period. During a seamless Phase I/II dose optimization study with simultaneous evaluation of toxicity and efficacy, that surgery could create an artificial efficacy response due to the reduction in tumor size after the surgery. Therefore, a surgery in this scenario could represent an intercurrent event.

The updates to the RECISTv1.1 [Eisenhauer et al., 2009] evaluation criteria specify that for a trial where surgery and radiotherapy are not part of the protocol treatment, "if a target lesion has been surgically removed or treated with radiotherapy prior meeting the response criteria, then the patient's response is not evaluable" thereafter [Schwartz et al., 2016]. Therefore, the information prior to the event of such surgery needs to be considered using a while-on-treatment strategy considering response to treatment up to the time of surgery. It is recommended to outline these implicit assumptions clearly and perform a sensitivity analysis to check their validity as required. The relevance of a while-on-treatment strategy might depend upon the type of estimand measure used. For example, if a while-on-treatment strategy implies omitting patients who have undergone surgery, this might omit informative data.

Though time-to-events outcome are less common in early phase dose optimization trials, missing data resulting from patients' premature withdrawal prior to surgery could undermine the robustness of the study results. One possibility is to consider efficacy and toxicities as delayed time-to-event outcomes, similar to the approach of Yuan and Yin [2009], where toxicity and efficacy are modeled as time-to-event outcomes that view the unobserved outcomes as censored events. Dropouts in such cases due to reasons related to the surgery result in informative censoring in trials. Sensitivity analyses using imputation methods are useful to examine the uncertainty due to informative censoring and address the robustness and strength of the study results. A supplementary analysis can be pre-specified in which censored outcomes due to the premature event-based study discontinuation are imputed based on adverse events that are clinically associated with the primary efficacy endpoint.

A composite variable strategy or a hypothetical strategy can instead be used in such cases to handle such intercurrent events. In the case of the composite variable strategy, the causes of the surgery would be required and the intercurrent event itself would carry information about the patient's efficacy outcome and would be considered as an efficacy response to treatment. For example, if the surgery was performed due to tumor shrinkage while still not meeting the response criteria, one might assume that the patient would have had at least a partial response and consider the response as favorable. If the surgery is undertaken due to clinician's choice and without prior evidence of tumor shrinkage, a while-on-treatment strategy is implementable considering the efficacy summary measure of patient's response prior to the surgery. If the surgery was undertaken due to external factors such as characteristic improvements, then a hypothetical strategy would be more suitable where the estimation of efficacy signal is performed assuming the hypothetical that the surgery was not performed. The implementation of the composite variable strategy and the hypothetical strategy in such a scenario for a Phase 1b trial handling Objective Response Rate (ORR) as the primary endpoint, is detailed in Englert et al. [2023] and is also applicable for evaluating efficacy outcomes in dose optimization studies implementing a seamless phase I/II strategy by simultaneously considering efficacy and toxicity. A case-by-case investigation is recommended for such an intercurrent event implementing different strategies.

### 3.7 Treatment Discontinuation Due to Disease Progression

FDA through its draft guidance [FDA, 2023] on single-arm trials to support accelerated approval, recommends the frequent use of response rate in oncology as the endpoint to support accelerated approval when the approval is based on data from single-arm trials. As per their guidance, appropriate criteria for assessing the response rate should be used. Response rates such as the disease control rate or the objective response rates does not consider an observation of a disease progression as a positive response and would therefore cause deterioration in the value of a response rate. However, it generally reflects an anticipated change in the underlying disease and treatment effect dynamics rather than an unexpected event or intervention and, accordingly, is not necessarily itself an intercurrent event. The primary objective of a dose optimization trial is to combine toxicity rates and clinical response rates into an utility measure to estimate the most desirable dose level that can be recommended for a phase II trial. Disease progression can be discussed as intercurrent events only as a pragmatic classification to aid study design as it often results in systematic withdrawal of patients from clinic visits. This can often affect the implied estimand and tends to require handling such discontinuation due to disease progression as an intercurrent event for the targeted estimand. For endpoints that require continuing clinic visits to observe, discontinuation due to disease progression will in practice generally need to be treated as intercurrent events when designing clinical trials [Siegel et al., 2024].

Patients with a global deterioration of health status requiring discontinuation of treatment without objective evidence of disease progression at that time is termed as 'symptomatic deterioration'[Eisenhauer et al., 2009]. Symptomatic deterioration is not considered as representative of an objective response. It is rather considered as a reason for stopping the treatment. An explicit hypothetical strategy could be considered in dealing with symptomatic deterioration envisaging a scenario by evaluation of target and non-target disease at the first observed time point.

For patients discontinuing the study for experiencing clinical progression or symptomatic progression, it is recommended that per protocol such patients continue to be scanned until radiologically defined progression [Englert et al., 2023]. After ensuring such follow-up visits, a treatment policy strategy can be implemented ignoring clinical or symptomatic progression. In practice, however, such follow-up visits may be hard to implement. A more realistic and practical approach therefore would be to implement the composite variable strategy if such follow-up visits may be inapplicable, where the discontinuation of treatment due to clinical progression prior to observing a response is included in the definition of an efficacy measure as a failure, which would reduce the desirability score accordingly. This is because this event of discontinuation can be considered informative on the outcome of the patient since documented progression may be expected soon. We argue that defining a principal stratum and focusing the analysis on patients that who are continuing in the trial would provide a biased approach towards estimating the target estimand for selecting the optimal dose level.

## 4 Comparing Different Estimands

Establishing the existence of an optimal biological dose is the key objective in dose optimization trials. Estimands in such trials should relate to the clinical question of interest that is being answered during the trial. This directly relates to the trial objectives that the target estimand focuses on. Different estimands are compared by examining how they define the target of estimation, particularly concerning intercurrent events and the population of interest. Key differences arise in how they handle the intercurrent events such as treatment interruptions, dose modifications, discontinuation and also the method of handling missing data. It might be seen that while comparing various estimands, a hypothetical or principal stratum estimand might produce more favorable effect estimates compared to other estimand strategies. However, to bring clarity in the interpretation of the utility function and the OBD, each estimand should be compared with the same estimate from the existing data. This aligns with the framework depicted in Fig 1. Deciding on an estimand while comparing various that produces a favourable effect should involve assessing which of the estimand best aligns with the specific clinical question and trial objectives An estimand strategy implemented for a specific trial objective should align with the same estimate from the existing data. Table 1 lists the various estimands for the intercurrent events (ICEs) that can occur in dose optimization trials and relates them to the respective targeted clinical question of interest. The estimate of the targeted estimand would be the maximum posterior expected utility function combining the patients' benefit-risk profile based on the data obtained applying a specific strategy for a clinical question of interest.

Table 1: Overview and Comparison of Possible Estimands

| Clinical Question | What is the utility score regardless of this ICE ? | What is the utility score considering this ICE as a failure event ? | What is the utility score, had this ICE not occurred ? | What is the utility score prior to this ICE? | What is the utility score, in the strata of participants not experiencing this ICE ? |
|---|---|---|---|---|---|
| **Estimands** | | | | | |
| Population | Defined through appropriate Inclusion/Exclusion in the trial protocol to reflect the target patient population for approval | | | | |

Table 1: Overview and Comparison of Possible Estimands (Continued)

| Variable /Endpoint | | Incidence of toxicity and response rate | | | | |
|---|---|---|---|---|---|---|
| Treatment Condition | | Investigational compound + any additional/ subsequent therapies | Investigational compound + any additional/ subsequent therapies | Investigational compound | Investigational compound | Investigational compound |
| Population-level Summary | | Utility Function ($U_j$ or $U_j^M$) | | | | |
| Estimation | | Expected Posterior Utility | | | | |
| ICE Handling Strategy | ICE: Treatment Discontinuation due to Toxicity | Treatment Policy | Composite Variable | Hypothetical | While on Treatment | Principal Stratum |
| | ICE: Use of Additional or Subsequent Therapy | Treatment Policy | Composite Variable | Hypothetical | While on Treatment | Principal Stratum |
| | ICE: Death | Treatment Policy | Composite Variable | Hypothetical | While on Treatment | Principal Stratum |
| | ICE: Surgery | Treatment Policy | Composite Variable | Hypothetical | While on Treatment | Principal Stratum |
| | ICE : Occurrence of Human ADAs | Treatment Policy | Composite Variable | Hypothetical | While on Treatment | Principal Stratum |

Table 1: Overview and Comparison of Possible Estimands (Continued)

| | | | | | |
|---|---|---|---|---|---|
| ICE: Discontinuation Due to Disease Progression | Treatment Policy | Composite Variable | Hypothetical | While on Treatment | Principal Stratum |
| ICE: Adherence to the Planned Treatment Regime | Treatment Policy | Composite Variable | Hypothetical | While on Treatment | Principal Stratum |

## 5　Considerations for Analyses Methods

The primary objective of a seamless Phase I/II dose optimization study with simultaneous evaluation of toxicity and efficacy is the identification of an optimal biological dose (OBD), if one exists. OBD here is defined as the admissible dose that has the highest desirability in terms of the risk-benefit trade-off. In the case of BOIN12, the admissible dose having the highest expected utility score is chosen at the end of the trial. If all doses in a trial are inadmissible, the trial is terminated. These trials are exploratory in nature and the methods of analyses are typically performed using summary measures. In contrast to the estimand addressing the questions "why" and "what" to estimate, described in the study protocol, the estimator fom the analysis method described in the statistical analysis plan addresses the question "how" to estimate the quantity of interest once the data have been collected. In dose optimization trials implementing a BOIN12 method, there are usually two quantities of interest: the OBD and the utility function at a given dose level. The former is the target of decision and is not directly estimated, while the latter is a target of estimation (hence, subject to estimand's definition). The link between these two quantities is a feedback control to determine which admissible dose would optimize the patients' desirability based on efficacy and toxicity information during the trial. Therefore, the estimated quantity is the utility function based on the mean expected posterior probability of a multinomial outcome based on efficacy and toxicity information in the trial and the OBD is, later on, derived from these estimated utility functions.

　Methods of statistical inference which are usually performed using Bayesian approaches such as the BOIN12 method, are based on the study design used at the design stage to identify the OBD. The final analyzed variable is summarized depending on the nature of the endpoint under consideration. Summaries for categorical variables include counts and percentages with one-sided or two-sided confidence intervals of these percentages. Pre-specified chosen strategies to handle intercurrent events determine how the numerator and denominator of the proportion of binary or multilevel outcomes will be calculated and/or included in the utility score calculations jointly summarizing the toxicity and efficacy information available. For example, denominators for proportions might consist of all treated subjects who have a baseline assessment and at least 1 adequate post-baseline response assessment or discontinued trial treatment prior to the first post-baseline response assessment or at least 1 postbaseline evaluable PK sample, as opposed to the "all randomized patients" populations commonly seen in traditional randomized clinical trials. Continuous variables are summarized using standard summary statistics (total number of patients, mean, standard deviation, median, minimum, and maximum) of the endpoint under consideration. Survival data such as the investigator assessed progression-free survival are often collected as a secondary endpoint to assess the preliminary antitumor activity of the compound in early phase dose optimization trials. Medians, as well as $25^{th}$ and $75^{th}$ percentiles (where evaluable), needs to be presented for such survival data. Where appropriate, 95% confidence intervals (CIs) around these Kaplan-Meier point estimates such as various percentiles of the survival curve are often presented. All treated subjects who have a baseline assessment and at least 1 adequate post-baseline response assessment or discontinued trial treatment prior to the first post-baseline response assessment or at least 1 postbaseline evaluable PK sample is most commonly used for such dose optimization trials implementing the BOIN12 design approach.

Patient-reported outcomes (PRO) provides a systematic and quantitative assessment of expected symptomatic side effects and their impact on function. FDA in its draft guidance [fda, 2024] mentioned that the inclusion of PROs should be considered to enhance the assessment of tolerability in dose optimization trials, as well as subsequent trials. Patient Reported Outcomes (PROs) provide patient-relevant information to support the safety and tolerability of the compound under investigation. The PRO-CTCAE, the patient-reported version of the clinician-reported Common Terminology Criteria for Adverse Events (CTCAE), are often used to assess symptomatic adverse events. To assess the overall impact of treatment side effect, a summary measure of the overall impact of treatment toxicity, based upon its association with the number and degree of adverse events in clinical trials can be used. The Functional Assessment of Chronic Illness Therapy - Item GP5 (FACT-GP5) are often used to assess the overall impact of such treatment side effects. Compliance rate are calculated at each time point as the number of completed assessments divided by the number of expected assessments, defined as the number of participants eligible to complete each instrument at that time. The selected PRO-CTCAE questions may be used in exposure-response analysis to explore response relationships with the dose to provide supportive safety information for dose optimization and dose selection. Descriptive statistics are usually reported for each PRO-CTCAE question and FACT-GP5 by each assessment timepoint and by dose. An all-data listing for all subjects should be presented in the final clinical study report.

Sensitivity analysis is important in clinical trials to explore the extent of robustness of the estimation associated with the primary estimand to deviations from the estimator's underlying assumptions. Following [Mercier F et al., 2022], sensitivity analyses is differentiated here from supplementary estimand analyses. As mentioned by Englert et al. [2023], any potential deviations from assumptions made for the primary estimator would need to explored using appropriate sensitivity analyses, such as testing missing not at random (MNAR) assumptions for missing data (originating, for example, from study discontinuations). Normally, assumptions about the MAR or MNAR are tested in the sensitivity analysis. However, when there is no control arm, the traditional control-based approaches such as pattern-mixture models to analyze the missing data will not be applicable. A tipping point analysis approach can be implemented by shifting the favorable response combining toxicity and efficacy into non-favorable and re-evaluating its effect on the predicted utility function. The shift value would represent the difference between the utility score of observed data and the missing data. If the shift value required to overturn the conclusion is so extreme that it is considered clinically implausible, this indicates robustness to the missing data assumptions. Here, the purpose would be to explore the plausibility of missing data assumptions under which the conclusion changes and to assess the robustness for later phase trials. Regulatory agencies often request such analysis as a sensitivity analysis under the Missing Not at Random (MNAR) assumption. If using a historical comparator, it is important also to verify that the historical data also used the same clinical question/estimands. While using model-based methods such as the EffTox [Thall and Cook, 2004] for identifying the OBD, it is important to check the validity of the model assumptions used to perform the dose optimization. This is because the identified estimand is heavily dependent on the assumption of the Bayesian model used at the design stage. In such a trial using a model-based method for dose optimization, the role of the trial statistician in the safety review committee becomes very important and the interim report at each stage by the safety review committee needs to cover the sensitivity analysis after fitting the Bayesian model to allocate the next cohort of patients to an optimum dose level and also at the final stage to identify the OBD based on the data obtained from the last cohort of patient. The sensitivity to the prior distribution used in the model can be checked at each interim stage by the designated trial statistician at the safety review committee, using a Haldane's prior [Haldane, 1932] and comparing it to the results obtained from the prior used for the model based design. A Gelman-Rubin [Gelman and Rubin, 1992] diagnostic test should also be performed to check for convergence of the Gibbs sampler method used for Bayesian inference at each stage to identify the optimum dose level.

When several strategies are deemed reasonable to address a specific intercurrent event, it is important to prioritize them according to the scientific question of interest. The most relevant strategy need to be used for the primary analysis to fulfil the primary objective of the trial. While the others should be used as the supplementary estimand analyses associated with different clinical objectives. Comparing the different results from the supplementary analyses clearly associated with the different targeted clinical questions of interest can give a deeper insight about the compound profile which can be used as an important information in further phases of the developmental process The choice of estimand is critical in causal inference, as different estimands can lead to different conclusions about the causal effect. For example, the treatment policy estimand estimates the causal effect of receiving the intervention, regardless of whether the intervention was actually received or not. Implementing multiple strategies for the intercurrent events leads to different estimands that bring in the complexity of interpreting the causal effect. The analysis methods need to be tailored according to the types of intercurrent events that can occur and also the order in which they can occur. While implementing a strategies for handling intercurrent events, it must be considered if known/unknown confounders predict both the intercurrent events and the outcome. The strategies to handle the intercurrent event needs to ensure that the causal effect is identifiable depending on the assumed data structure. While combining multiple hypothetical strategies, one approach is to use the estimation methods such as inverse probability of missingness weighting to model the occurrence of the different intercurrent event types being handled by the hypothetical strategy separately. . The critical point of consideration is the clinical question of interest and the trial objective that relates to the clinical question of interest. The handling strategies of the intercurrent events must be aligned with the trial objectives to answer the clinical question of interest.

The inferential method to estimate the OBD at the end of the clinical trial depends on the trial design (e.g., Ji3 + 3, EffTox, BOIN12, PRINTE, etc.) implemented for the dose optimization process. However, the analysis method for the secondary and exploratory endpoints is unrelated to the dose optimization design method implemented in the trial and depends on the respective objectives and endpoints of the trial and the data collected during the trial process. As per FDA [fda, 2024], relevant nonclinical and clinical data (such as PK, PD, safety, tolerability, dosage convenience, and activity), as well as the dose- and exposure-response relationships should be evaluated to select the optimum dose to move ahead in the developmental process. The exposure–response (E-R) analysis can therefore be broadly utilized as a supportive analysis towards the findings from the implemented design and to provide a better characterization of OBD. FDA in their draft guidance [fda, 2024] emphasized that for every dose optimization trial, the dose- and exposure-response relationships should be evaluated to select a dosage(s) to be recommended as the phase II dose. The summary measures for exposure are usually area under the curve (AUC) and/or maximum concentration (Cmax), instead of the concentrations over time. The response variables include safety endpoints and efficacy endpoints, which typically are continuous, categorical or time-to-event data. In most cases, E-R analysis is performed on the use of simple regression models via a nonlinear least square method in order to evaluate relationships between exposure and response at a single time-point. Details of E-R analysis have been discussed in Overgaard et al. [2015]. The dose optimization analysis is commonly conducted in the end of the trial once the PK, efficacy and safety data are available.

## 6  Real-Life Case Study for Estimand Example

**Table 2**  Attributes of the Estimand and Strategies to Handle the ICEs

| Attributes of Estimand | Determining the OBD |
|---|---|
| Trial Objective | Identifying the OBD |
| Estimand | Utility function to Identify the Optimal Biological Dose |
| Target Population | Adults with metastatic/ unresectable solid tumors, with PD since last therapy |
| Endpoint | Binary endpoint combining efficacy and toxicity (DLT) rate to obtain the OBD |
| Treatment under Consideration | NM21-1480. The treatment period starts at the first dose of study drug. |
| Population Level Summary | Dose desirability measured by the utility function of efficacy and toxicity rates |
| ICE (Treatment Discontinuation due to Toxicity) | Composite Variable Strategy, handled as failure. |
| ICE (Death) | Composite Variable Strategy, handled as failure. |
| ICE (Use of Additional/Subsequent Therapy) | While-on-treatment Strategy, data before event is included to determine outcome |
| ICE (Discontinuation due to Disease Progression) | Composite Strategy, discontinuation considered as a failure |
| ICE (Occurrence of ADAs) | Treatment Policy Strategy, data before and after the event is included. |
| ICE (Switching Dose Levels) | Treatment Policy Strategy, data before and after the event is included. |
| ICE (Surgery by Clinician's Choice) | While-on-treatment Strategy, data before event is included. |
| ICE (Surgery representing treatment failure) | Composite Variable Strategy |
| ICE (Surgery conducted by external factors) | Hypothetical Strategy |

The case study presented in this manuscript, which was designed and conducted by Numab Therapeutics AG, has been described and the biological results of the study have been published elsewhere [Luke et al., 2022]. This is a first-in-human, multicenter, open-label, phase1/2a trial of the NM21-1480 compound in advanced solid tumors. The primary objective was to estimate MTD, based on the number of dose-limiting toxicities (DLTs) observed at a specific dose level. A BOIN design was used and consisted eight planned escalating dose levels; dose levels 1 to 8 in the range of 0.15mg-1400mg.

In summary, 26 patients with various primary solid tumors were enrolled in the study. Of the 26 enrolled, 23 were evaluable for efficacy.  One patient experienced a DLT. In the 8mg-800mg dose range, disease control, i.e., at least stable disease at first assessment at 8 weeks, occurred in 13/23 patients (54%).  PD activity remained stable at a broad dose range (24mg- 800mg).

After enrolling into dose level 1 (0.15 mg), subsequent dose levels were only opened if the previous dose level was deemed well tolerated. The first dose level was to enroll a minimum of one patient as per the accelerated titration approach. If a Grade 2 or higher adverse event (AE) was observed during the evaluation period or when dose level 5 was reached, a minimum of three patients were to be enrolled per dose level in accordance with the BOIN design dosing rules. Cohorts of three patients were recruited after the first dose level enrolling three patients.

The drug was administered as a single intravenous (IV) infusion approximately every 14 days for a total of two infusions per treatment cycle. The DLT observation period was therefore 28 days.

The target toxicity rate for the MTD was set as ≤30% (i.e, $\phi$ =0.3), and the maximum sample size was 27 with a maximum of 12 patients per cohort. If the observed DLT rate at the current dose was ≤0.236, the next cohort of patients would be treated at the next higher dose level; if it was ≥ 0.359, the next cohort of patients would be treated at the next lower dose level. If the DLT rate was > 0.236 or < 0.359 then the next cohort of patients would be treated at the current dose. A 1-patient-per-dose dose-escalation process was applied until the first ≥Grade 2 toxicity was observed or dose level 5 (80 mg dose level) reached.

While enrolling patients into dose levels 1 to 5, it became clear that the drug had a benign safety profile. Furthermore, newly available pharmacodynamic (PD) data suggested that toxicity may not increase monotonically with dose and that PD activity might plateau due to the affinity-balanced design of the molecule, i.e., activity might initially increase and then plateau over a relatively broad dose range before decreasing. The biological explanation for such a bell-shaped dose-response relationship is that at high concentrations the target engagers for the drug may become saturated resulting in "insulating effects" that restrict drug activity. Based on the PD and emerging clinical data it was decided to amend the study design to remove the highest pre-specified dose level from the dose escalation scheme [Luke et al., 2022].

Clark T et al. [2024] investigated the application of BOIN12 to the same study setting and compared the BOIN and the BOIN12 methods through extensive simulations pointing out the merits of the later approach in such settings. They showed using the simulation tools in the BOIN suite [Zhou et al., 2019], how targeting the OBD instead of the MTD would have been a better choice in this setting. The authors illustrated using the the real data that the BOIN design targeting the MTD determines the dose level 7, 800 mg, as the optimal dose. Whereas, using the BOIN12 design targeting the OBD determines the dose level 4, 24 mg, as the optimal biological dose. However, the authors did not use an estimand framework while demonstrating their findings that can account for the intercurrent events occurring during the trial. In this manuscript we demonstrate how the proposed estimand framework can be incorporated for this case study in case of implementing the BOIN12 method to identify the OBD. Therefore, the primary objective of the study implementing the BOIN12 method would be to identify the OBD that is to be estimated as a combination of the toxicity and the efficacy in the trial. Implementing the proposed estimand framework would enable informed decision making and bring clarity in the descriptions of the benefits and risks for interpreting the OBD and selecting the recommended phase II dose while implementing the BOIN12 approach.

In the case of a benign toxicity compound such as NM21-1480 of the present case study, any treatment discontinuations due to adverse events constitute treatment failures and need to be accounted by the composite variable strategy . Similarly, any event of death in this study during the escalation process need to be considered as a terminal event related to the treatment under consideration and should be dealt with using the composite variable strategy. For patients taking additional anti-cancer therapies during the escalation process, a while-on-treatment strategy need to be implemented here considering the responses prior to the use of the anti-cancer therapy, as opposed to a composite variable strategy that would have considered subsequent therapy as a response failure outcome. This is because, taking additional anti-cancer therapies, though may be assumed to be associated with a non-favourable response, is not considered detrimental to the standard of prognosis for the patients in terms of treatment efficacy or safety. Therefore, such an event is not considered as a treatment failure and the while-on-treatment strategy would be most appropriate in such scenarios where the focus is to identify the OBD based mainly on the effect of the compound of interest. As in usual dose optimization trials in oncology, if the intention-to-treat principle is used for analyzing the primary estimand of the study, an intercurrent event of dose switching in this trial should be handled by the treatment policy strategy. If the solid tumour becomes operable for the patient in the trial during the treatment period, a while-on-treatment strategy can be used if the surgery is undertaken due to clinician's choice and without evidence of prior tumor shrinkage. Using RECIST v1.1 [Eisenhauer et al., 2009] as a measure of efficacy evaluation in the trial for estimating the dose-response relationship, the composite variable strategy can be implemented for the intercurrent event of discontinuation of patients due to disease progression. As the trial aims to further develop the compound NM21-1480 in advanced solid tumors, a treatment policy strategy would need to be used in the event of any occurrence of ADA. Therefore, it is evident that all estimand strategies to handle the intercurrent events are associated with the compound under investigation, its trial objective and the clinical question of interest. Dose optimization trials should adequately evaluate a range of dosage(s) and select the dosages to be further investigated based on all available clinical data and intercurrent events, as well as preliminary understanding of dose- and exposure-response. This would enhance the quality of decision making of selecting the OBD at the end of the trial and enhance the interpretation of the utility function used to identify the OBD. Table 2 provides a tabulated summary of the attributes of each considered estimand together with details on the strategies to address the ICEs.

For the analysis of the primary estimand in this trial, the frequency of DLTs and that of the complete and partial responses will be tabulated by dose for patients in the trial and all efficacy and toxicity data needs to be listed by dose. The population summary measure of the utility function, combining both toxicity and efficacy information, would inform decisions related to dose-level desirability that is calculated using the quasi-beta-binomial [Lin et al., 2020b] model that converts the desirability calculations from the utility function into a beta-binomial modeling.

## 7 Discussion

The ICH E9 (R1) addendum ich [2020] primarily focuses on randomized controlled trials. Noticing this, Englert et al. [2023] discussed an estimand framework for single arm Phase 1b dose expansion trials aimed at exploring early efficacy signals of the compound under investigation. Although briefly mentioning FDA's Project Optimus, the authors focused mainly on Phase 1b studies to characterize or refine the dose-response relationship in an MTD setting without considering a seamless dose escalation process considering efficacy and toxicity measures simultaneously, as is needed in the OBD case. In addition to toxicities of various grades, efficacy is a key parameter for seamless Phase I/II dose optimization studies with simultaneous evaluation of toxicity and efficacy. Expanding from the Phase 1b dose expansion trial case or the Phase 2 single arm trial case, this manuscript discusses an estimand framework for early phase oncology trials designed to identify OBD as the primary estimand, balancing safety and efficacy during the dose escalation process, and outlining how these considerations can be considered while selecting intercurrent event handling strategies and lead to different considerations than for MTD trials. Since early phase dose optimization trials targeting OBD are becoming essential for sponsors in their Investigational New Drug (IND) application process, this estimation framework can be applicable to such studies.

Estimands help sponsors make internal decisions by clearly characterizing how the objective of the clinical trial can be served with respect to what is to be estimated in the trial. For seamless Phase I/II dose optimization studies, as both the toxicity and efficacy information are considered along with PK and PD data to characterize the dose and exposure response relationships in such early phase trials, the applicability of the estimand framework can be expanded and considered for these cases. This manuscript discusses an estimand framework which a sponsor could employ while preparing their IND as well as while running dose optimization trials targeting the OBD.

In contrast to late clinical development where treatment comparison is the key objective, early phase clinical trials focus on exploration. However, oncology drug development is particularly challenging with low rates of approval for novel treatments when compared to other therapeutic areas [Jaki et al., 2023]. The FDA therefore promotes innovation and modern technology to overcome corresponding challenges. As in other exploratory trials, it may not always be possible to prospectively define all the intercurrent events in the study protocol. For example, some intercurrent events, like the occurrence of anti-drug-antibodies, might now be anticipated and thus any pre-specification of a related strategy might not be feasible. Strategies to handle intercurrent events may also depend on the risk–benefit trade-off considered during the escalation process of the trial. For rare intercurrent events, it may not be possible to align on a suitable strategy while developing the protocol, and monitoring the occurrence of some events on a case-by-case basis may become necessary. In general, study teams in these kinds of early phase trials should implement a transparent approach related to the clinical question of interest to be answered by the trial.

Early phase dose optimization trials should include a sound PK sampling plan in order that the PK exposure parameters can be appropriately estimated. This is critical because the PK exposure parameters play important role for identification with the optimized dose, such as when (PK) exposure-response modeling analysis is applied. Of note, in case the PK sampling is sparse and insufficient to calculate the PK exposure such as AUC and Cmax, a population PK model can be developed to simulate the required PK variables. In addition, population PK modeling can also be useful in assessing effects of covariates and special populations. On the other hand, PK outcomes are usually available much faster than clinical response or toxicity outcomes during the trial process. The model-assisted BOIN12 design has been recently extended to PKBOIN12 design [Sun and Tu, 2024] by integrating PK information into both the dose-finding algorithm and the final OBD determination process. The authors in Sun and Tu [2024] also extended their methodology to address the challenges of late-onset toxicity and efficacy outcomes. However, an estimand framework has not been discussed for such dose optimization trials integrating PK information from the dose and exposure-response models into the dose optimization design to find the OBD. Therefore, the present work can be extended to suggest an estimate framework for trials incorporating PK information into the dose optimization trial design, such as the PKBOIN12 design [Sun and Tu, 2024] to properly inform decision making by having a structured framework that strengthens the collaboration between various disciplines involved in the execution of such trials. One limitation of the approaches presented in this paper is that intra-patient escalation and extrapolation beyond studied dose levels cannot be performed and is viewed as further work to be examined in the future. Mechanistic PK-PD models allow the integration of prior information and totality of data to be used for model development. This mechanistic model is usually applied to predict different efficacy and toxicity profiles at different dose levels which provides the opportunity to conduct dose optimization exercises.

For next generation anti-cancer therapies such as immunotherapy, reliance on short-term toxicity data such as DLT, is not sufficient to estimate the optimized dose. Delayed clinically adverse events in early phase immunotherapy trials have been reported. Also low-grade toxicities e.g., Grade 1 diarrhea is not uncommon. These toxicities impact the patients' quality of life significantly especially when lasting long time. In addition patient report outcomes (PROs) might also provide valuable information about safety, tolerability and efficacy on top of the observations from clinicians. All these information can be considered and integrated into utility functions for optimized dose selection.

In a real-life clinical trial, a dose with a desirable risk-benefit trade-off may not be among the competing doses being studied. The described BOIN12 [Lin et al., 2020b] approach does not consider dose extrapolation beyond the studied doses. One approach that can be studied further is to develop this framework for dose optimization design methods that consider dose extrapolation using data from the previous doses or from pre-clinical safety assessment study. This can be done using a model-based design approach such as the EffTox design [Thall and Cook, 2004] where a fitted statistical model can borrow information across dose levels to extrapolate beyond the studied dose level, if the dose with desirable risk-benefit tradeoff is not be among the ones studied in the trial. The common model-based approaches such as the EffTox design or the model-assisted approaches such as the BOIN12 design are based on the conventional interpatient dose escalation strategy, where each patient can be treated only at one dose level during the escalation process. Patients can however not be recycled for higher dose levels when efficacy response is observed but no dose limiting toxicity is observed. There are situations when patients can receive higher doses if the administered dose is well tolerated. When the accrual rate is slow, one might also implement approaches such as the AIDE method [Zhou et al., 2023] using the patient's individual safety data, as well as other enrolled patient's safety data to consider intra-patient dose escalation to identify the maximum tolerated dose (MTD). However, a design approach to identify the OBD that considers intra-patient dose escalation is not known to the best of our knowledge and can be developed as a future research work.

As the dose-optimization trials in oncology are exploratory, the strategies as outlined in ICH E9 (R1) may need further development for unplanned intercurrent events for first-in-human studies. This would provide more flexibility to seamless Phase I/II trials to facilitate drug and biologic development. Appropriately justified and transparent strategies implemented in dose-optimization oncology trials, along with a suitable study design method, would ensure transparency in communication and reporting of trial results and thus enhance the efficiency and approval rate of novel oncology treatments. One important area of development of estimand framework for exploratory trials in oncology is when late-onset toxicities are a serious concern in phase I trials, or when long efficacy observation windows significantly enhances the trial duration for dose optimization. Though the authors have briefly discussed the trials with long efficacy observation windows and the study design methods available for such trials, the strategies considered in this paper primarily assumes that dose-limiting toxicities are assessed across the first cycle of therapy and that the efficacy observation window is not large enough to delay the trial duration significantly. Such assumptions are more common for hematological studies. If a compound causes late-onset toxicities, however, an undesirably large number of patients may be treated at toxic doses before any toxicities are observed. For example, the cisplatin trial in pancreatic cancer in Muler et al. [2004] had an assessment window of 9 weeks and thus such delayed outcomes needs to be accounted for while building the model for targeting the OBD. When efficacy observation windows are large, one option to handle it is to introduce some PD biomarkers (eg: percentage decrease in Interleukin-2 receptor alpha chain (CD25)) that can be considered as a surrogate for measuring efficacy in the dose optimization trials. However, a suitable estimand strategy needs to be developed while considering such PD biomarkers as a surrogate measure for efficacy endpoints that have a long observation window. A transparent estimand framework handling the intercurrent events with suitable strategies in such trials with delayed efficacy and toxicity outcomes would ensure clarity in the understanding of the targeted estimand and enhance the efficiency of the design and conduct of the trial.

### Acknowledgments

This work is a result of a collaboration between the Population Health Sciences Department of Newcastle University and the Early Phase Oncology Working Group (EPOWG) within IQVIA. The authors would like to specially thank Newcastle University and IQVIA for providing resources in support of completing this research work. Special thanks also goes to Numab Therapeutics AG for their case-study which has been used here to develop the framework. Special thanks are also to the IQVIA colleagues: Jin Chen, Michael O' Kelly and Jeffrey Hodge for their thorough reviews and feedback. The authors would like to thank the anonymous reviewers and the special issue editor for their comments that led to a substantially improved version of the article.

### Author contributions

Ayon Mukherjee (AM), Jonathan Moscovici (JM) and Zheng Liu (ZL) initiated the project. The concept originated from AM and then further developed by JM and ZL. The case-study was initiated by AM in collaboration with Numab Therapeutics AG. All authors contributed to the manuscript drafting and read and approved the final manuscript.

### Financial disclosure

This research work has been supported by Newcastle University and IQVIA. The authors declare that there have been no additional funding sources for this research work.

### Conflict of interest

The authors declare no potential conflict of interests.

### Data Availability Statement

Data sharing not applicable to this article as no datasets were generated or analysed during the current study.

### ETHICS APPROVAL STATEMENT

Not Applicable

### PATIENT CONSENT STATEMENT

Not Applicable

### CLINICAL TRIAL REGISTRATION

Not Applicable